\documentclass{aastex631}
\shorttitle{M-T relation for clusters, groups, and galaxies}
\shortauthors{Iu. V. Babyk and B. R. McNamara}

\begin{document}

\title{The Halo Mass-Temperature Relation for Clusters, Groups, and Galaxies}

\author{Iurii V. Babyk}
\affiliation{Center for Astrophysics $|$ Harvard-Smithsonian, 60 Garden Street, Cambridge, MA, 02138, USA}

\affiliation{Main Astronomical Observatory of the National Academy of Sciences of Ukraine, 27 Academica Zabolotnoho str., Kyiv, 03134, Ukraine}

\author{Brian R. McNamara}
\affiliation{Department of Physics and Astronomy, University of Waterloo, 200 University Avenue West, Waterloo, ON, N2L 3G1, Canada}

\begin{abstract}

The halo mass-temperature relation for a sample of 216 galaxy clusters, groups, and individual galaxies observed by $Chandra$ X-ray Observatory is presented. Using accurate spectral measurements of their hot atmospheres, we derive the $M-T$ relation for systems with temperatures ranging between 0.4-15.0 keV. We measure the total mass of clusters, groups, and galaxies at radius $R_{2500}$, finding that the $M_{2500} \propto T^{\alpha}$ relation follows a power-law with $\alpha$ = 1.65$\pm$0.06. Our relation agrees with recent lensing studies of the $M-T$ relation at $R_{200}$ and is consistent with self-similar theoretical prediction and recent simulations. This agreement indicates that the $M-T$ relation is weakly affected by non-gravitational heating processes. Using lensing masses within $R_{200}$ we find $M_{200}-T$ follows a power-law with slope 1.61$\pm$0.19, consistent with the $M_{2500}-T$ relation. No evidence for a break or slope change is found in either relation. Potential biases associated with sample selection, evolution and the assumption of hydrostatic equilibrium that may affect the scaling are examined.  No significant impacts attributable to these biases are found.  Non-cool-core clusters and early spirals produce higher scatter in the $M-T$ relation than cool-core clusters and elliptical galaxies.

\end{abstract}


\section{Introduction} \label{sec:intro}
Although hierarchical structure formation is well understood, the influence of cooling and heating on the intracluster (ICM), intragroup (IGM) and interstellar media (ISM) of clusters, groups, and galaxies is not. This makes them interesting because of the effects regarding the galaxy-cluster scaling relations and for the formation of galaxies used in numerous cosmological tests \citep{Voit05}. According to the $\Lambda$CDM theory, clusters, groups, and galaxies formed similarly \citep{Kaiser:86, Evrard:02}. This means that the physical characteristics of cosmic structures, such as temperature, luminosity, and mass should scale as power-laws with forms $L_X \propto T^{2}$ and $T \propto M^{2/3}$ \citep{Mushotzky:84, Kaiser:86, Mushotzky:97, Kaiser:91, Evrard:02, OSullivan_sample:03, Borgani:04, Kravtsov:06, Nagai:07, Giodini:13}.

However, recent X-ray observations show significant deviations from the predicted self-similarity, showing $L_X \propto T^{2.7-3.0}$ and $M \propto T^{1.6-1.8}$ \citep{Vikhlinin:06, Vikhlinin:09, Babyk:12jphst, Babyk:12a, Babyk:12d, Babyk:13a, Babyk:12oap, Babyk:13oap, Babyk:14a, Babyk:18aasp}. Moreover, these deviations increase with decreasing mass-scale of cosmic structure \citep{Boroson:10, Kim:13, Kim:15, Babyk:17scal}. Even a small disagreement between theory and observations indicates that current theoretical models are not accurate. Disagreements in self-similarity predictions are likely related to non-gravitational processes that occur at the cores of clusters, groups, and galaxies, such as AGN feedback, shock heating, supernova heating, etc. \citep{Ma:13, Pulatova:15, Choi:15, Babyk:16, Main:17, Babyk:17ent}. Although some progress has been obtained both in observation and theory related to quantifying and understanding X-ray scaling relations \citep{Markevitch:98, Lin:04, Ohara:06, Babyk:14a, Babyk_book:15}, it is still unclear how the non-gravitational processes affect different mass-scales.

In contrast with the luminosity-temperature relation, the $M-T$ scaling relation is tighter.  Therefore,  atmospheric temperature is apparently a better proxy for halo mass than X-ray luminosity. The mass of cosmic structures is widely used to define the cosmological parameters and mass function of the Universe \citep{Finoguenov:01, Voit:05, Morandi:07, Vikhlinin:09, Ettori:13}. By extending studies of X-ray scaling relations from clusters to low-mass, faint groups and individual galaxies we are able to further probe the long-term impact of non-gravitational heating processes as well as to explore the galaxy evolution. Here we focus on the study of such impact to explore the $M-T$ relation in a wide range of temperature- and mass-scale systems. 

Previous measurements of the  $M-T$ relation are usually obtained for galaxy clusters and bright groups \citep{Vikhlinin:09, Babyk:14a}. It was found that the $M-T$ scaling relation of nearby clusters is less affected by non-gravitational processes than other scaling relations, showing smaller departures from self-similarity. However, these departures vary with mass, increasing from the most relaxed, massive galaxy clusters to individual galaxies. The $M-T$ relation for the most relaxed galaxy clusters follows a power-law with a slope of 1.6-1.8 \citep{Nevalainen:00, Tozzi:00, Finoguenov:01, Vikhlinin:09, Babyk:13a, Babyk:14a}, while the mass of low-mass systems scales with temperature with a slope of 1.8-2.2 \citep{Boroson:10, Babyk:17scal}. The simultaneous fitting of $M-T$ relation for clusters, groups, and galaxies has been performed for only small samples \citep{Finoguenov:01, Vikhlinin:09, Lieu:16}, and mostly for systems with a temperature $\gtrsim$ 2 keV. 

Adopting hydrostatic masses, \citet{Arnaud:05} measured the $M-T$ relation for a sample of ten nearby galaxy clusters in the $2-9$ keV energy band and over a range of density contrasts. They found that the $M-T$ relation for the systems within $\rm M_{500}$ and $kT > $3.5 keV is consistent with  self-similarity. For the whole sample, they got steepening 1.71$\pm$0.09. 

\citet{Kettula:13} and \citet{Lieu:16} used weak-lensing masses to measure the $M-T$ relation. Kettula combined 10 galaxy groups in the COSMOS field with 55 galaxy clusters from the literature. Their slope of 1.48$\pm$0.13 is consistent with the self-similar expectations. They found that group masses based on hydrostatic equilibrium are lower than expected from the self-similarity indicating a high bias that reaches 30-50\% in 1 keV atmospheres. 

In contrast with the \citet{Kettula:13} result, \citet{Lieu:16} found a steeper slope (1.67$\pm$0.12) for their sample of 96 systems. They explored the scaling between the weak-lensing mass and X-ray temperature. The large scatter ($\sigma$ = 0.41) of their fit is driven by disturbed galaxy clusters.

\citet{Farahi:18} used dynamical mass measurements in the XMM-XXL survey to explore $M-T$ and other scaling relations. They found an even steeper slope (1.89$\pm$0.15) for the $M-T$ relation compared to slopes based on the X-ray and weak-lensing mass measurements. They also found no significant redshift evolution. 

\begin{table*}
\caption{Low-scale systems sample taken from Babyk et al. (2023) in prep.}\label{tab1}
\centering
\begin{tabular}{lccccccccccc}
\hline
 && \\
Name      & R.A. & Decl. & ObsIDs & Exposure & Type & BCG  &  $z$     &  $D_{\rm A}$ &  $D_{\rm L}$  &  $N_{\rm H}$ \\
          &  (J2000)   & (J2000)    &       &    ks     &      &         &          &   Mpc  &  Mpc    &  10$^{20}$ cm$^{2}$ \\
          & (2) & (3) & (4) & (5) & (6) & (7) &(8) & (9) & (10) & (11) \\
&& \\
\hline
&&\\
3C449   & 22:31:20.625 & +39:21:29.81 & 11737, 13123 & 53.0, 60.7 & E & $\surd$ & 0.017085 & 71.255 & 73.7 & 12.2 \\
HCG62   & 12:53:05.6 & -09:12:21.0 & 10462, 10874 & 68.0, 52.0 & E & & 0.01472 & 61.615 & 63.4 & 3.01 \\
IC1860  & 02:49:33.878 & -31:11:21.94 & 10537 & 37.7 & E4 & $\surd$ & 0.0229 & 94.658 & 99.0 & 2.07 \\
IC4765  & 18:47:18.150 & -63:19:52.14 & 15637 & 15.0 & E4 & $\surd$ & 0.015034 & 62.76 & 64.7 & 7.11 \\
NGC1132  & 02:52:51.82 & -01:16:29.0 &  3576 & 40.1 & E &         & 0.023133 & 95.587 & 100.1 &  5.22  \\
NGC1521  & 04:08:18.937 & -21:03:06.98 & 10539  & 50.0 & E & & 0.01415 & 59.281 & 61.0 & 2.54  \\
NGC1600  & 04:31:39.858 & -05:05:09.97 & 4283, 4371  & 27.1, 27.1 & E3 &   & 0.0156 & 65.267 & 67.3 &  4.71 \\
NGC2300  & 07:32:20.486 & +85:42:31.90 & 4968  & 46.1 & E & & 0.006354 & 26.944 & 27.3 & 5.48  \\
NGC2305  & 06:48:37.295 & -64:16:24.05 &  10549  & 10.0  & E   &   & 0.011671 &  49.083 & 50.2  & 6.87 \\
NGC4406  & 12:26:11.814 & +12:56:45.49 & 16967  & 20.0  & E &    & 0.000747 & 3.195 & 3.2 & 2.58 \\
NGC4486  & 12:30:49.423 & +12:23:28.04 & 5826, 5827  & 128.4, 158.2 & E &      & 0.004283 &  18.221 & 18.4 & 2.54 \\
NGC4936 &  13:04:17.091 & -30:31:34.71 & 4997, 4998   & 14.1, 15.1   & E &     & 0.010397 &  43.812 & 44.7  & 5.91 \\
NGC5129 & 13:24:10.001 & +13:58:35.19 & 6944, 7325 & 21.1, 26.1 & E  &         & 0.022969 & 94.933 & 99.3  & 1.76 \\
NGC5419 &  14:03:38.771 & -33:58:42.20 & 4999, 5000 & 14.7, 15.0 & S0 & $\surd$ & 0.013763 & 57.695 & 59.3  & 2.36 \\
NGC6407 & 17:44:57.664 & -60:44:23.28 & 5896  & 21.8   & S0  & & 0.015427 & 64.504 & 66.5  & 7.69 \\
NGC6868 & 20:09:54.082 & -48:22:46.25 & 11753  & 73.5 & E &    & 0.00952 & 40.171 & 40.9  &  4.94 \\
NGC7619 & 23:20:14.524 & +08:12:22.63 &  2074, 3955 & 27.0, 37.9  & E2 &   & 0.012549 & 52.704 & 54.0 &  4.92 \\
NGC7796 & 23:58:59.910 & -55:27:30.12 & 7061, 7401  & 53.9, 20.2 & E &   & 0.011221 & 47.224 & 48.3 &  2.22 \\
\hline
\end{tabular}
\end{table*}

Here we present the $M-T$ relation for 216 $Chandra$ galaxy clusters, bright and faint groups of galaxies as well as individual galaxies (mostly early-type galaxies). We use our previous measurements of the temperature and total mass obtained recently in \citet{Hogan:17a, Pulido:17, Babyk:17prof}, Babyk et al. 2023, in prep. and \citet{Main:17} (hereafter Paper I, II, III, IV, V). Due to the lack of the physical radius of individual galaxies and groups, we use $R_{2500}$\footnote{$R_{2500}$ is the radius within which the mean density is 2500 times the critical density.} scaled radius to measure the total mass at this radius for all objects. Although our sample is not homogeneous and statistically incomplete, it is the largest to date for which individual measurements of X-ray temperature and total mass of low-mass systems at $R_{2500}$ have been performed. We also use previously published lensing measurements of mass at $R_{200}$ to explore possible evidence of break reported in previous studies of $M-T$ relation for much smaller samples.

A standard $\Lambda$CDM cosmology with the following parameters: $H_0$ = 70 km s$^{-1}$ Mpc$^{-1}$, $\Omega_{\Lambda}$ = 0.7, $\Omega_{M}$ = 0.3 has been assumed. The errors are given for 1$\sigma$ confidence level unless otherwise mentioned. 

\section{Data collection}\label{sec:2}

Our sample consists of 55 clusters and groups analyzed in Paper I, 55 cool-core clusters and groups analyzed in Paper II, 40 groups and individual galaxies presented in Paper III, 18 individual galaxies analyzed in Paper IV, and 48 clusters and groups in Paper V. The main physical characteristics of the 18 individual systems analyzed in Paper IV are shown in Tab.~\ref{tab1}. The analysis of X-ray observations was performed similarly for all five data sets. Here the data reduction is given briefly whilst a detailed description is provided in refereed papers. 

The observations with at least 10~ks of exposure time were downloaded from the HEASARC\footnote{http://heasarc.gsfc.nasa.gov/} archive. The data reduction was performed using {\sc ciao} (version 4.3 was used in Paper I, II, V and 4.8 in Paper III, IV) software package and the calibration files, {\sc caldb} (version 4.2.1 was used in Paper I, II, V and 4.8.1 in Paper III, IV). The pipeline processing, re-processing, re-screening, time-dependent gain correction, and others have been performed using \texttt{xpipe, chandra\_repro}, and \texttt{acis\_process\_events} tools. 

The black-sky background files each observation and the \texttt{lc\_clean} routine provided by M. Markevitch to search the background flares were used. We combined the multiple observations and perform reprojection into an observation with higher exposure time. The background-subtracted images were created and then cleaned using the \texttt{wavdetect} task.

We divided the background-subtracted images on concentric annuli and centered them at the maximum of X-ray emission. Since the number of photons in the X-ray observations of clusters and galaxies are different, the total number of X-ray photons in each annulus was different as well. In the case of galaxy clusters, we plotted annuli with at least 3000 photons, while the annuli of groups and galaxies consisted of at least 700 photons. We extracted the source background spectra individually for each observation and created the {\sc arf} and {\sc rmf} files using the \texttt{mkwarf} and \texttt{mkacisrmf} tools. Our spectra were grouped with at least 20 counts/bin using \texttt{grppha} routine. 

Two different spectral models to fit our spectra were selected. Both of them included a thermal component to define the X-ray parameters of hot gas. In addition to the thermal component, the X-ray spectra of low-mass systems have been fitted with non-thermal components to define the contribution of low-mass X-ray binaries and other stellar sources to the total X-ray emission. The {\sc phabs*(apec+po+mekal+po)} and {\sc phabs*mekal} models were applied to fit the low-mass system spectra and the spectra of galaxy clusters respectively. We used the {\sc xspec} environment to perform our spectral fitting. The detailed information of spectral fitting, including values of fixed and non-fixed parameters, is given in Papers I, II, III, IV, and V. The deprojection routine was achieved to define more accurate spatially resolved profiles \citep{Babyk:18oap}. We applied the same spectral models as the projected spectra and used the deprojected profiles for our following analysis.

Temperature and normalization profiles were used to plot spatially resolved electron density profiles. The total mass profiles were derived using hydrostatic equilibrium and spherical symmetry assumptions. The total mass profiles of galaxy clusters and groups in Papers I, II, and V were found assuming the Navarro-Frank-White dark matter density profile. The equation of hydrostatic mass was used to derive the total mass profiles of 58 low-mass systems in Papers III and IV. 

\section{Results}\label{sec_res}
We use our density profiles to define $R_{2500}$ by integrating them until the mean mass density is 2500 times the critical density at a given cosmology. The 2500 overdensity level is less useful than 500 due to the stronger effect of the bright central region on the temperature and density measurements. However, the 2500 level is a powerful tracer for objects in a wide range of total mass. The physical radius of low-mass systems is about 10-20 kpc while the characteristic radii $R_{2500}$ and $R_{500}$ are about 200 and 600 kpc, respectively. The extrapolation to $R_{2500}$ is much more accurate for low-mass systems than those to $R_{500}$ or $R_{200}$. To define the total mass of low-mass systems we extrapolate their mass profiles to $R_{2500}$ using the linear slope of the last 5 points in log-log space. It is straightforward to scale the mass of CDM halos to any overdensity level \citep{Hu:03}.
\subsection{$kT$ and $M_{2500}$ vs. $z$}

\begin{figure}
\centering
\includegraphics[width=0.49\textwidth]{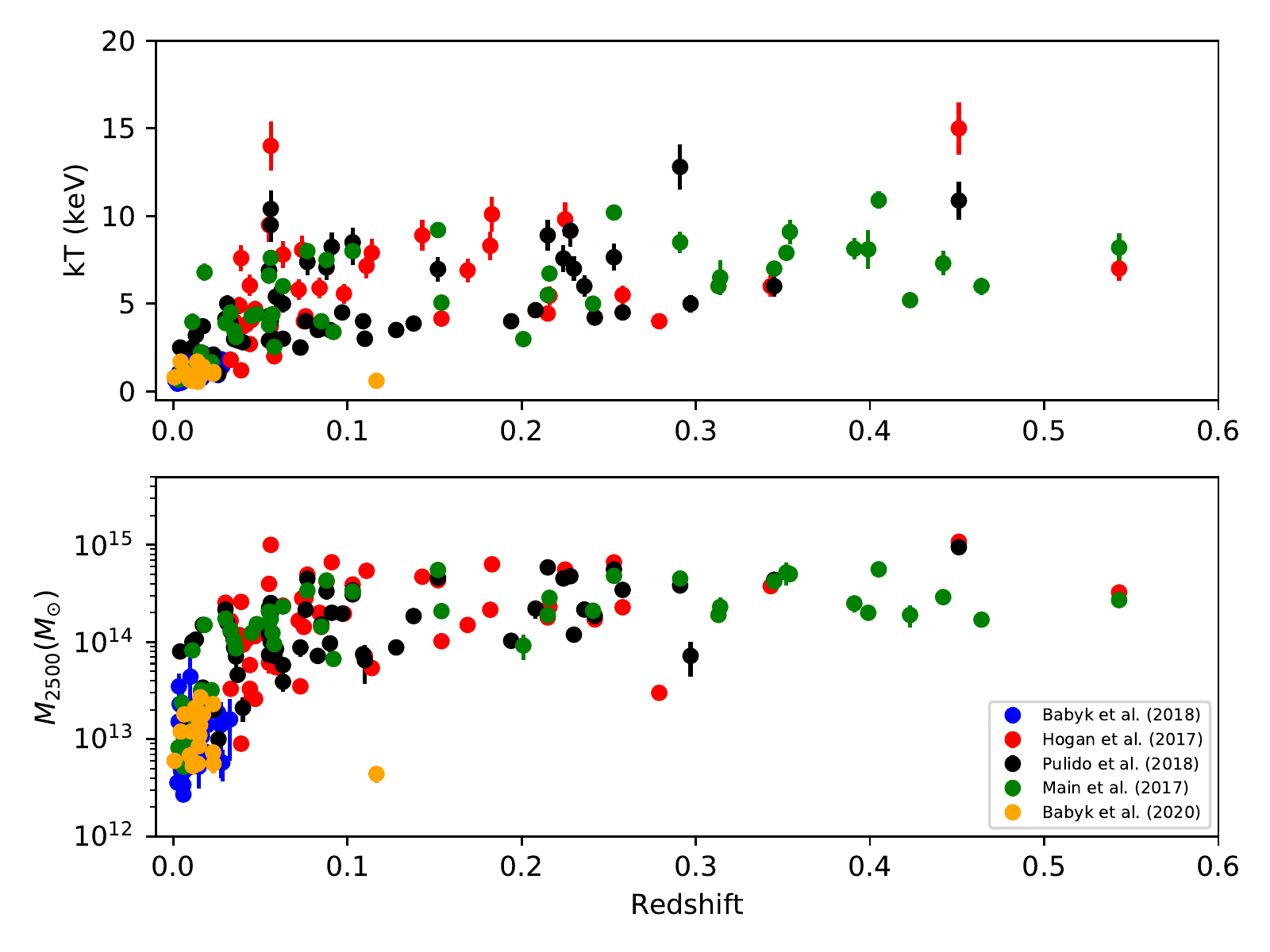}
\caption{The derived temperature and total mass at $R_{2500}$ plotted against the redshift of our systems.}\label{fig_11}
\end{figure}

The redshift distribution of this sample spans $\approx$ 0.0007-0.7. Temperatures and total masses are plotted against redshift to explore dependencies in Fig.~\ref{fig_11}.  Systems beyond redshift $>$0.1 have similar temperatures and masses, while nearby systems ($z<$0.1) show a significant dependence of temperature and mass on redshift. This result agrees with previous work (see, for example, \citet{Mantz:16, Lieu:16, Farahi:18}), and suggests that a redshift evolution term should be considered in the $M-T$ scaling relation.

\subsection{$M_{2500}-T$ relation}
\begin{figure}
\centering
\includegraphics[width=0.49\textwidth]{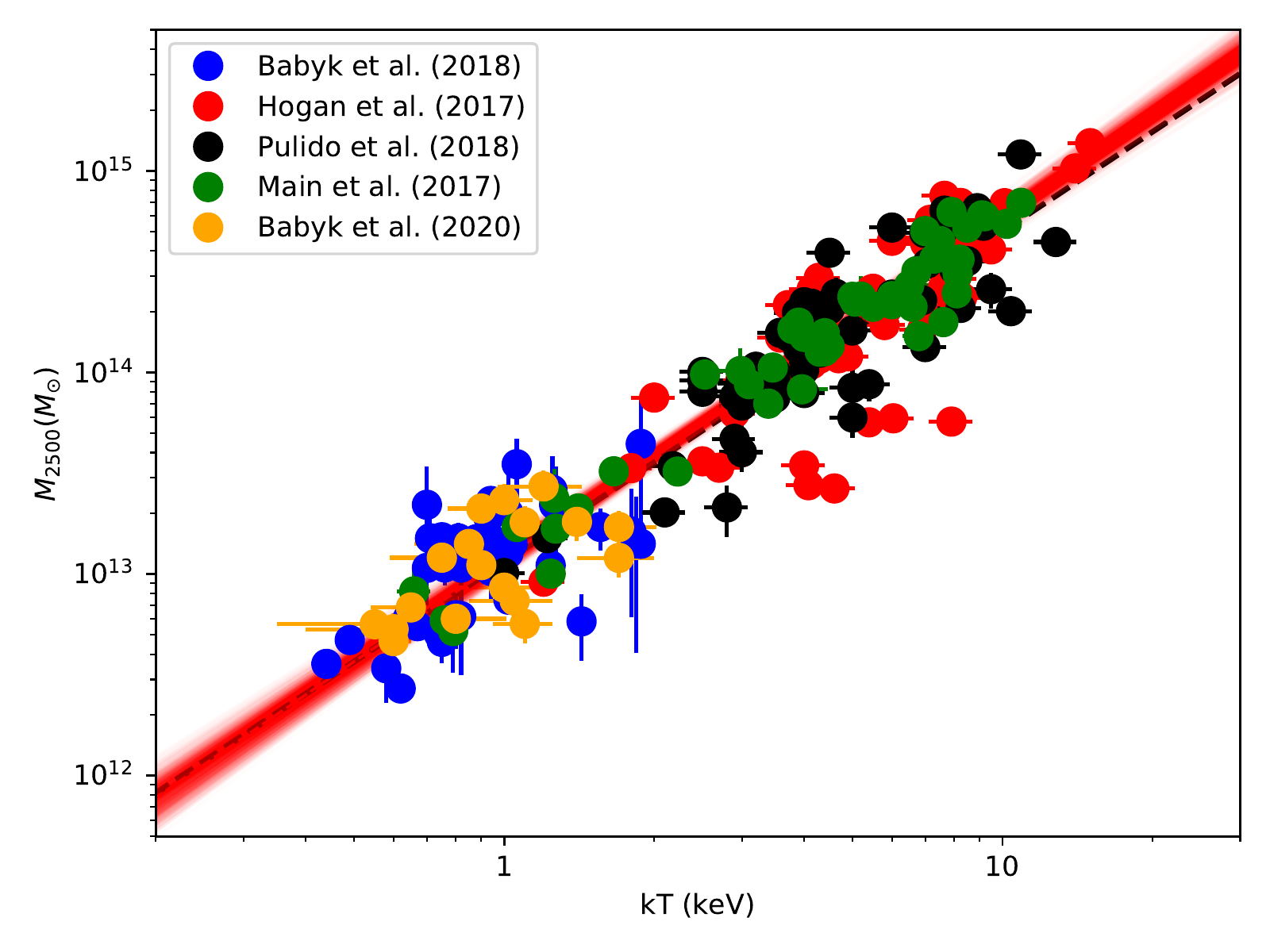}
\caption{The derived total mass at $R_{2500}$ plotted against the average temperature of our systems. The red lines represent Kelly's best-fitting model with a standard deviation. The dashed line is given for comparison with the result obtained by \citet{Vikhlinin:06} for a sample of 13 systems in the range of 0.7-12.0 keV in temperature and total mass derived at $R_{2500}$.}\label{fig_1}
\end{figure}

In Fig.~\ref{fig_1}, masses within $R_{2500}$ are plotted against X-ray gas-weighted temperatures. The red and black points represent data taken from Papers I and II, respectively. The blue and yellow points represent data obtained in Papers III and IV (Babyk et al. 2023, in prep.), while the green points represent data from Paper V. The self-similar evolution of the $M-T$ relation is expected to follow $M \propto T^{1.5}$. Redshift dependence is modeled by fitting our relation with and without the redshift factor, discussed above. The observed relation was fitted by a single power-law model, $log(y)=a+b\cdot log(x)$. 

The likelihood-based approach of \citet{Kelly:07} was applied to fit the $M-T$ relation. Kelly's regression method adopts a Bayesian likelihood function with an intrinsic scatter,  implemented in the Python-based Linmix package\footnote{https://github.com/jmeyers314/linmix.}. Uncertainties in the best-fit parameters were estimated by running 15,000 iterations of a Markov Chain Monte Carlo procedure.
Kelly's method assumes Gaussian-distributed residuals.  The Shapiro-Wilks (SW) and Anderson-Darling (AD) tests were applied to check the normality of their residuals. We find $p >$ 0.9 in our $M-T$ relation, confirming that the residuals are indeed Gaussian distributed. The Spearman and Pearson correlation tests were applied to measure the significance of the relationship between mass and temperature. 
Results of our calculations, including best fits, uncertainties, $p$-values for null-hypothesis, and normality with variance provided by Kelly's algorithm are given in Table~\ref{tab_res}. 

\begin{table*}
\centering
\caption{$M-T$ scaling relation of the form log($y$) = $a$ + $b$ log($x$).}\label{tab_res}
\begin{tabular}{cllccccccc}
\hline
& Sample & $N$ &  $a$ & $b$ & $p$-Pearson & $p$-Spearman & $p$-AD & $p$-SW & Scatter \\
\hline
& $M_{2500}-T$ (BCES) & 216 &  13.12$\pm$0.02 & 1.62$\pm$0.09 & $>>$0.0001 & $>>$0.0001 & 0.97 & 1.11 & - \\
& $M_{2500}-T$  & 216 &  13.08$\pm$0.04 & 1.65$\pm$0.06 & $>>$0.0001 & $>>$0.0001 & 0.97 & 1.61 & 34$\pm$8\% \\
& $M_{2500}(z)-T$ & 216 & 13.07$\pm$0.03 & 1.69$\pm$0.05 & $>>$ 0.0001 & $>>$0.0001 & 0.97 & 1.64 & 33$\pm$9\% \\
X-ray & Clusters  & 145 & 12.96$\pm$0.10 & 1.78$\pm$0.15 & $>>$0.0001 & $>>$0.0001 & 0.97 & 1.25 & 35$\pm$11\% \\
& Groups \& Galaxies  & 71  & 13.08$\pm$0.05 & 1.74$\pm$0.32 & $>>$0.0001 & $>>$0.0001 & 0.97 & 1.41 & 29$\pm8$\% \\
& Cool-Core (CC) & 122 & 12.88$\pm$0.12 & 1.72$\pm$0.13 & $>>$0.0001 & $>>$0.0001 & 0.97 & 1.22 & 30$\pm$6\%\\
& Non-Cool-Core & 23 & 11.78$\pm$0.67 & 1.81$\pm$0.75 & 0.011 & 0.021 & 0.47 & 0.67 & 42$\pm$14\%\\
& CC+Groups \& Galaxies & 193 & 13.10$\pm$0.03 & 1.66$\pm$0.04 & $>>$0.0001 & $>>$0.0001 & 0.98 & 1.55 & 31$\pm$6\%\\
& Groups & 15 & 13.11$\pm$0.10 & 1.73$\pm$0.13 & $>>$0.0001 & $>>$0.0001 & 0.96 & 1.44 & 25$\pm$7\%\\
& Ellipticals & 46 & 13.05$\pm$0.07 & 1.70$\pm$0.11 & $>>$0.0001 & $>>$0.0001 & 0.95 & 1.48 & 23$\pm$4\%\\
& Lenticulars+Spirals & 10 & 12.76$\pm$0.17 & 1.85$\pm$0.22 & 0.028 & 0.031 & 0.42 & 0.54 & 39$\pm$11\%\\
& CC+Ellipticals+Groups & 183 & 13.08$\pm$0.05 & 1.67$\pm$0.03 & $>>$0.0001 & $>>$0.0001 & 0.98 & 1.58 & 30$\pm$5\%\\
\hline
& $M_{200}-T$ & 75 & 13.47$\pm$0.07 & 1.61$\pm$0.19 & $>>$0.0001 & $>>$0.0001 & 0.93 & 1.45 & 46$\pm$13\% \\
Lensing & Clusters & 30 & 13.51$\pm$0.10 & 1.65$\pm$0.31 & $>>$0.0001 & $>>$0.0001 & 0.94 & 1.22 & 23$\pm$9\% \\
& Groups and Galaxies & 45 & 13.56$\pm$0.11 & 2.10$\pm$0.47 & $>>$0.0001 & $>>$0.0001 & 0.93 & 1.31 & 27$\pm$12\% \\
\hline
\end{tabular}
\end{table*}

The best-fit slope and normalization of the $M-T$ relation without the redshift term are $b$ = 1.65$\pm$0.06 and $a$ = 13.08$\pm$0.04, respectively. A redshift evolution term,  $M_{2500}E(z)$, where $E(z)$ = $\sqrt{\Omega_M(1+z)^3+\Omega_{\Lambda}}$, was explored. No significant change or improvement was found in slope or other parameters indicating the effects of redshift evolution are negligible over our redshift range. The lack of change is because our sample is dominated by low redshift objects, as shown in Fig \ref{fig_11}. We, therefore, omit the evolution term in further analysis.

\subsection{Literature Comparison}
Here we compare our results with previous work. Our slope is slightly flatter than found by \citet{Nevalainen:00, Tozzi:00, Shimizu:03} but it agrees with \citet{Allen:01, Finoguenov:01, Vikhlinin:06, Zhang:08, Sun:09}. The difference between our results and those of the first set of authors may be due to different measurement techniques and calibration with the previous generation of X-ray observatories (e.g., $ASCA$, $ROSAT$).

Our $M-T$ relation is tighter for massive clusters than for low-mass systems, due in part to the uncertainty associated with the extrapolation of the total mass to $R_{2500}$. We find that the $M-T$ relation for our galaxy clusters (kT $>$ 2 keV) follows $b$ = 1.78$\pm$0.15 and $a$ = 12.97$\pm$0.11 power-law form. Our results are consistent within uncertainties with the slopes of $M-T$ relations of \citet{Vikhlinin:06, Echmiller:11} for the 2500 and 500 overdensity levels.  A comparison between our parameters and those of earlier studies is given in Tab.~\ref{tab:my_label}.  Ours is the largest sample available, particularly at low masses, of individual temperature and hydrostatic mass measurements out to $M_{2500}$. 

Our slope is slightly steeper in value than those obtained in numerical simulations performed by \citet{Borgani:04, Nagai:07, Borgani:11}, but well within the uncertainties. \citet{Borgani:04} performed large cosmological hydrodynamical simulations using the $TREE+SPH$ code $GADGET$ to simulate the X-ray observable properties of galaxy clusters and groups in a wide range of temperatures. The influence of AGN feedback on the M-T and other scaling relations was recently shown in numerous papers including \citet{Pellegrini2018, LeBrun2014, Planelles:15, Pike2014, Li:15} and many others. These simulations showed that AGN feedback can successfully suppress
cooling in cosmological simulations and isolated systems. \citet{Sijacki:06, Chon:2016, Rosito:2021} also found that the AGN feedback is able to break the self-similarity. The contribution of AGN and supernovae feedback, stellar mass loss, thermal condition, and/or cosmic rays varies, depending on the mass of the source, increasing from the clusters to low mass systems. However, the AGN feedback is the largest source in the hot atmospheres for all masses \citep{McNamara:07, McNamara:12, Babyk:17scal}. Thus, the discrepancy between theory and observations of the M-T and other scaling relations can be attributed to AGN feedback.


\begin{table*}
    \centering
    \caption{Comparison with previous results.}
    \begin{tabular}{ccc|cccc}
    \hline
    & Our results & & & Previous results \\
    Relation & \# & Data & Relation & \# & Data & Ref \\
    \hline
    &&& $M_{2500}-T^{1.51\pm0.27}$ & 6 & X & \citet{Allen:01} \\
    &&& $M_{2500}-T^{1.64\pm0.06}$ & 13 & X & \citet{Vikhlinin:06}\\
    &&& $M_{500}-T^{1.61\pm0.11}$ & 13 & X & \citet{Vikhlinin:06} \\
    &&& $M_{500}-T^{1.65\pm0.26}$ & 37 & X & \citet{Zhang:08} \\ 
    &&& $M_{vir}-T^{1.9}$ & & X & \citet{Shimizu:03} \\
    $M_{2500}-T^{1.65\pm0.06}$ & 216 & X & $M_{500}-T^{1.64\pm0.04}$ & 88 & X & \citet{Finoguenov:01}\\
    &&& $M_{500}-T^{1.65\pm0.04}$ & 57 & X & \citet{Sun:09}\\
    &&& $M_{500}-T^{1.75\pm0.06}$ & 174 & X & \citet{Echmiller:11}\\
    &&& $M_{500}-T^{1.53\pm0.08}$ & & X & \citet{Vikhlinin:09}\\
    &&& $M_{500}-T^{1.59\pm0.05}$ & & S & \citet{Borgani:04} \\
    &&& $M_{2500}-T^{1.55\pm0.05}$ & & S & \citet{Borgani:04}\\
    \hline
    &&& $M_{500}-T^{1.42\pm0.19}$ & 52 & L & \citet{Mahdavi:13}\\
    $M_{200}-T^{1.61\pm0.19}$  & 75  & Lensing & $M_{500}-T^{1.78\pm0.37}$ & 38 & L & \citet{Lieu:16} \\
    &&& $M_{500}-T^{1.48\pm0.13}$ & & L & \citet{Kettula:13} \\
    &&& $M_{200}-T^{1.89\pm0.15}$ & & L & \citet{Farahi:18}\\ 
    \hline
    \end{tabular}
    \label{tab:my_label}
\end{table*}
Our normalization is higher than those obtained in previous X-ray analyses based on $ROSAT$ or $ASCA$ \citep{Nevalainen:00, Tozzi:00}, but agrees with $XMM-Newton$ \citep{Arnaud:05} and $Chandra$ \citep{Vikhlinin:06} measurements. \citet{Vikhlinin:06} found (1.25$\pm$0.05)$\times$10$^{14}$ for $\Delta$ = 2500, while we found (1.51$\pm$0.10)$\times$10$^{14}$. 

Our normalization is lower by 25\% than those obtained numerically by \citet{Borgani:04}.  However,  we are consistent with high-resolution simulations by \citet{Nagai:07, Borgani:11} that include star formation, radiative cooling, and feedback from supernovae. \citet{Nagai:07} showed that X-ray mass estimates within $R_{500}$ are lower than those obtained in simulations. \citet{Rasia:05} suggested the 30\% higher normalization of the $M-T$ relation found by \citet{Borgani:04} is due to their adoption of spectral temperatures while \citet{Rasia:05} measurements are emission-weighted temperatures.

\subsection{Selection bias}
Here we discuss the potential biases that can affect the slope, normalization, and variances in our $M-T$ scaling relation.

Our sample includes early-type galaxies (elliptical and lenticular), spiral galaxies, brightest clusters, group galaxies, and clusters with cool and non-cool cores. When the low-mass systems are concerned, fossil groups (one bright central galaxy with an X-ray luminance counterpart) and similar halo masses with multiple galaxies but faint in X-ray (non-fossil groups) produce a bias which can have an impact on the estimated best-fit parameters and scatter.

Second, the degree of relaxation can introduce bias. Our Galaxy clusters include mostly relaxed cool-core systems while the low-mass systems are more likely unrelaxed. Moreover, the limited radial extent out to which emission in the low-mass system can be traced adds uncertainty and potential bias.  X-ray emission from  ellipticals
and groups is detectable to fairly large radii. But their lower temperature atmospheres, $\sim$0.1-1 keV, can be difficult to distinguish from the Galactic foreground. In addition, our X-ray analysis is limited by the field of view of $Chandra$ instruments.

In the case of the full sample, the best-fitting parameters are 1.65$\pm$0.06 for slope and 34$\pm$8\% for the scatter. We performed individual fittings for clusters, groups, and galaxies (see Tab.~\ref{tab_res}) and found that for all three sub-samples the best-fit slopes are similar within uncertainties with the slope obtained for the full sample. The scatter showed similar agreement within errors as well.

The largest scatter in the cluster sub-sample is produced by non-cool core (NCC) clusters. This is seen in the  \citet{Pulido:17} and \citet{Hogan:17a} samples (black and red points respectively). The \citet{Pulido:17} sample consists of cool-core clusters only, showing a lower scatter (29$\pm$5\%) than those given by red points (38$\pm$9\%) taken from \citet{Hogan:17a} sample. However, we found that this selection bias has no measurable impact on the slope, which is similar for the cool-core and non-cool-core sub-samples.  Fits excluding NCC clusters were performed and no significant differences in the slope or normalization were found. However, the scatter decreased from 34$\pm$8\% to 31$\pm$6\%.  The same issue related to the scatter in CC and NCC clusters was obtained by \citet{Lieu:16}, based on lensing mass measurements. 

Similar fits were performed for the low-mass sub-samples, ellipticals, early spirals+spirals, and groups.  Again, the slopes were similar between sub-samples but the scatter about the mean differed. The fit only for ellipticals gives 23$\pm$4\% scatter which is $\sim$ 30\% lower than for a full low-mass sub-sample. The highest scatter is produced by spirals, 39$\pm$11\% which may be unrelaxed systems where hydrostatic equilibrium may be violated (see the bias related to hydrostatic equilibrium below). The group sub-sample shows a similar slope and scatter to ellipticals. Similar results were found by \citet{Babyk:19}. 


A potential bias related to the log-log extrapolation of total mass profiles to $R_{2500}$ was explored. $M_{2500}$ was estimated by extrapolating beyond the outer five points of the total mass profile 
measured to $R_{2500}$.  In this instance, the slope and normalization of the best-fitting profile don't change, while the scatter of the $M_{2500}-T$ relation has increased by a few percent, 37$\pm$9\%. 

\subsection{Other biases}

Here we focus on other systematic biases. Perhaps the most important potential bias is the assumption of hydrostatic equilibrium. That hydrostatic mass estimates are biased compared to other methods is well documented (see \citet{Rasia:06, Nagai:07, Meneghetti:10, Smith:16}). To explore this bias X-ray mass estimates at $R_{2500}$ 
are compared to weak-lensing masses at the same radius for 25 galaxies from \citet{Harris:17} and 8 clusters from \citet{Okabe:16}. Both \citet{Harris:17} and \citet{Okabe:16} control systematic biases in their mass estimates. 

We define $\alpha$ as the geometric mean ratio of the hydrostatic mass to the weak-lensing mass for a sample of $N$ targets
\begin{equation}
    \alpha = exp \left[ \frac{\sum_{i=1}^{N}w_i \ln(M_{X,2500}/M_{WL,2500})}{\sum_{i=1}^{N}w_i} \right],
\end{equation}{}
where $w_i$ is the assigned weight of each target. The uncertainty on $\alpha$ is estimated as the standard deviation of the geometric means of 10000 bootstrap samples each numbering $N$ targets.

\begin{figure}
\centering
\includegraphics[width=0.49\textwidth]{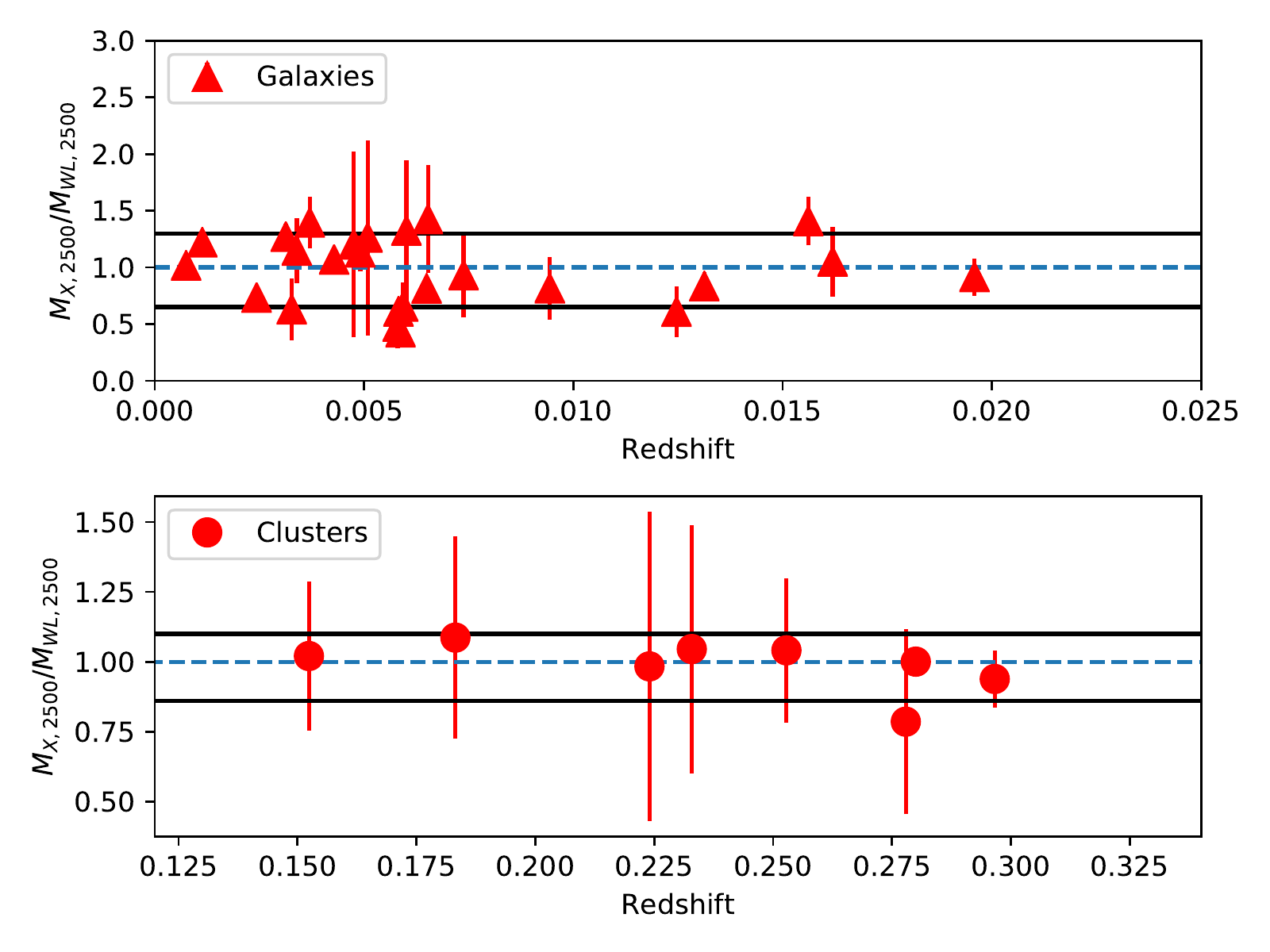}
\caption{Ratio of X-ray-based hydrostatic mass to weak-lensing mass versus redshift for galaxies (top panel) and clusters (bottom panel). The dashed lines correspond unity while the solid lines show the $\pm$3$\sigma$ confidence interval.} \label{fig_he}
\end{figure}

We find $\alpha$=0.94$\pm$0.08 for galaxies and $\alpha$=0.96$\pm$0.04 for clusters (errors for both slopes are on the mean). With clusters and galaxies combined, $\alpha$=0.94$\pm$0.09. \citet{Martino:14} found $\approx$0.93 and \citet{Smith:16} got 0.95, in accordance with our results. The ratio of X-ray masses to lensing masses is plotted for both galaxies and clusters against redshift in Fig.~\ref{fig_he}.  No statistically significant trend is seen in both cases. Summarizing: our X-ray and lensing masses agree with their uncertainties for both clusters and galaxies.

\citet{Kettula:13, Kettula:15} pointed out that hydrostatic equilibrium may be compromised by AGN feedback.  Thus, masses based on hydrostatic equilibrium may be biased, and underestimated, for low-mass systems. No strong bias is found here. 

\section{$M-T$ relation based on lensing mass measurements}

To explore the $M-T$ relation based on the lensing data we use previously published lensing mass estimates at $R_{200}$ for both galaxies and clusters. Lensing masses for 38 systems were found in \citet{Harris:17} for galaxies and \citet{Hoekstra:07, Kubo:09, Hoekstra:13} for clusters. Lensing masses for other clusters, groups, and galaxies with X-ray temperatures were included for a total of 75 systems. The lensing targets expanded the temperature range of our $M-T$ relation to 0.1-15.0 keV. The $M_{200}-T$ relation was fitted similarly to $M_{2500}-T$. Best-fit slope, $\alpha$ = 1.61$\pm$0.19, and normalization, 13.47$\pm$0.07, are consistent with each other (see Tab.~\ref{tab_res} for more details). 

\begin{figure}
\centering
\includegraphics[width=0.49\textwidth]{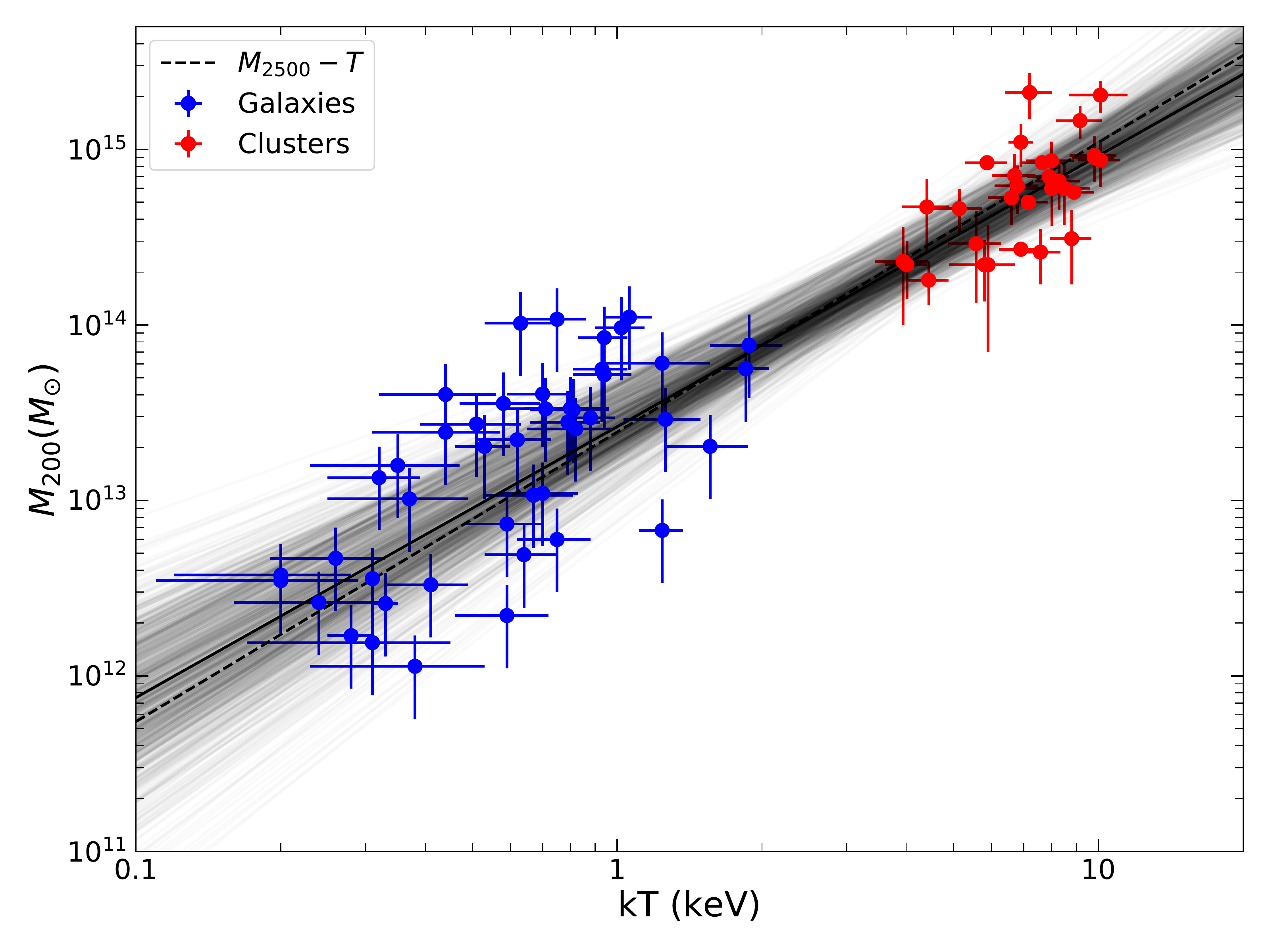}
\caption{Lensing mass at $R_{200}$ plotted against the average temperature of our sample. The solid line and shaded area represent Kelly's power-law fitting. The dashed line corresponds to the best-fit slope of X-ray $M_{2500}-T$ relation.} \label{fig_2}
\end{figure}

No evidence for a break or slope change in the $M_{200}-T$ relation is seen. Our results show that the $M_{200}-T$ relation has a slope of 2.1$\pm$0.5 within 0.1-2.0 keV and a slope of 1.7$\pm$0.3 within 2.0-15.0 keV which is less than 1$\sigma$ difference. We show the $M_{200}-T$ in Fig.~\ref{fig_2} by plotting 75 lensing systems in blue and red which represent galaxies/groups and clusters respectively. 

The $M_{200}-T$ relation is consistent with our $M_{2500}-T$ relation (black dashed line in Fig.~\ref{fig_2}) and self-similarity, but with large uncertainties. We also checked another potential bias in the M-T relation which is related to the impact of mass assembly history. Cosmological simulations of \citet{Chen:2019} showed that halos that have gained mass more recently are likely to have lower temperatures, as their gas is still not completely thermalized. Thus, we should see the difference in the slope of the M-T relation measured between 200 and 2500. However, we found no such difference.


Several studies have predicted and obtained a systematic difference between the physical characteristics of low-mass systems such as groups and/or galaxies compared to galaxy clusters, for example,  a steepening of the $L_X-T$ relation \citep{Ponman:96, Xue:00, Dave:08, Babyk:17scal}, or a flattening in the $L_X-\sigma$ relation \citep{Mahdavi:00, Mahdavi:13}. 
Similarly to other studies, we also found that all best-fit parameters and intrinsic scatter are larger for low-mass systems, while the slope is consistent within errors with the cluster fit. 
However, the best-fit normalization of the $M-T$ relation for low-mass systems is lower as well compared to the clusters, and this could cause the possible steepening. 

No break as expected with a broken power-law model is found, although a slight steepening at the low-mass end is evident.  The data, however, cannot rule-out a break due to the large scatter at the low-mass end and the dearth of data between 2-4 keV where a break was reported in previous studies \citep{Xu:01, Finoguenov:01, Afshordi:02, delpopolo:02, Echmiller:11, Capelo:12, Lovisari:15, Hammami:17}. The observed steepening at the low-mass end in the M-T relation as well as in the L-T and L-M relations are evident in \citet{Babyk:17scal}. The steepening is likely the imprint of AGN feedback, which would have a larger impact on lower mass atmospheres.
The power law relationship $M_{2500}-T$ is likely continuous over all masses as no slope change is found. 
Recently, \citet{Arnaud:05} analyzed a sample of massive clusters and showed that the slope of the $M-T$ relation is stable at all the overdensities. This agrees with our results. \citet{Lovisari:15} verified whether or not this is also true at the group scale by fitting the relation at $R_{2500}$ and $R_{1000}$ as well. They found that the slope is quite stable: 1.61$\pm$0.07 at $R_{2500}$, 1.71$\pm$0.13 at $R_{1000}$, and 1.65$\pm$0.11 at $R_{500}$. Recent lensing studies of the $M-T$ scaling relation for both groups ($T \approx$ 1-3 keV) and clusters showed a slope in good agreement with self-similarity \citep{Kettula:13}. 

The $R_{2500}$ relationship is tighter than the $R_{200}$ relation, in part because of the large extrapolation required to evaluate $R_{200}$. We have assumed that the mass profiled for all systems is similar, which may not be true. 

\section{Conclusions}\label{sec_summary}
The mass-temperature scaling relation is presented using a sample of galaxy clusters, groups, and individual galaxies observed by $Chandra$ X-ray Observatory. A total of 216 targets, including 71 low-temperature ($<$ 2 keV) systems were studied. This sample includes twice the number of low-mass systems compared to earlier studies. A tight relationship between temperature and mass is found.  Our results are summarized as follows.
\begin{itemize}
\item Using spatially resolved mass and temperature profiles, the $M_{2500}-T$ scaling relation over the temperature range of 0.4-15.0 keV and the mass range of 10$^{13}$-10$^{15} M_{\odot}$ is found. With the addition of weak lensing masses, the temperature range was extended to 0.1 keV.
\item The observed relation is slightly steeper than the self-similar theoretical prediction, 1.65$\pm$0.06 slope with a scatter of 34$\pm$8\%. 
\item Selection bias, hydrostatic equilibrium bias, and other sources of systematic error were explored. The largest scatter about the mean relationship is found for non-cool core clusters at the high-mass end, and early spirals and spirals at the low-mass end of $M-T$ scaling relation. These potential biases have little impact on the slope and normalization.
\item The derived slope and normalization of $M-T$ relation is consistent with numerical simulations that include the effects of cooling, star formation, and feedback.  The $M-T$ relation obtained here using X-ray techniques to measure mass profiles is broadly consistent with gravitational lensing. 
\item No evidence for a break in the power-law relationship between 2 keV and 4 keV is found in the  $M_{2500}-T$ and $M_{200}-T$ relationships. 
\end{itemize}


\begin{acknowledgments}

IB acknowledges financial support from the research grant for laboratories/groups of young scientists of the National Academy of Sciences of Ukraine in 2021-22 and from XMM-Newton Grant Number 80NSSC19K1056. BRM acknowledges support from the Natural Sciences and Engineering Research Council of Canada. This research has made use of data obtained from the Chandra Data Archive and the Chandra Source Catalog, and \software{CIAO \citep{Fruscione:06}, Sherpa \citep{Freeman:2001, Doe:2007, Burke:2020}, ChiPS \citep{Germain:2006}, XSPEC \citep{Arnaud:96}, linmix (https://github.com/jmeyers314/linmix)}. We thank all the staff members involved in the Chandra project. 
\end{acknowledgments}

\bibliography{paper}

\begin{thebibliography}{}
\expandafter\ifx\csname natexlab\endcsname\relax\def\natexlab#1{#1}\fi
\providecommand{\url}[1]{\href{#1}{#1}}
\providecommand{\dodoi}[1]{doi:~\href{http://doi.org/#1}{\nolinkurl{#1}}}
\providecommand{\doeprint}[1]{\href{http://ascl.net/#1}{\nolinkurl{http://ascl.net/#1}}}
\providecommand{\doarXiv}[1]{\href{https://arxiv.org/abs/#1}{\nolinkurl{https://arxiv.org/abs/#1}}}

\bibitem[{{Afshordi} \& {Cen}(2002)}]{Afshordi:02}
{Afshordi}, N., \& {Cen}, R. 2002, \apj, 564, 669, \dodoi{10.1086/324282}

\bibitem[{{Allen} {et~al.}(2001){Allen}, {Schmidt}, \& {Fabian}}]{Allen:01}
{Allen}, S.~W., {Schmidt}, R.~W., \& {Fabian}, A.~C. 2001, \mnras, 328, L37,
  \dodoi{10.1046/j.1365-8711.2001.05079.x}

\bibitem[{{Arnaud}(1996)}]{Arnaud:96}
{Arnaud}, K.~A. 1996, in Astronomical Society of the Pacific Conference Series,
  Vol. 101, Astronomical Data Analysis Software and Systems V, ed. G.~H.
  {Jacoby} \& J.~{Barnes}, 17

\bibitem[{{Arnaud} {et~al.}(2005){Arnaud}, {Pointecouteau}, \&
  {Pratt}}]{Arnaud:05}
{Arnaud}, M., {Pointecouteau}, E., \& {Pratt}, G.~W. 2005, \aap, 441, 893,
  \dodoi{10.1051/0004-6361:20052856}

\bibitem[{{Babyk}(2012{\natexlab{a}})}]{Babyk:12jphst}
{Babyk}, I. 2012{\natexlab{a}}, Journal of Physical Studies, 16, 1904

\bibitem[{{Babyk}(2016)}]{Babyk:16}
---. 2016, Astronomy Reports, 60, 542, \dodoi{10.1134/S1063772916040028}

\bibitem[{{Babyk} {et~al.}(2012){Babyk}, {Melnyk}, \& {Elyiv}}]{Babyk:12d}
{Babyk}, I., {Melnyk}, O., \& {Elyiv}, A. 2012, Advances in Astronomy and Space
  Physics, 2, 56

\bibitem[{{Babyk} \& {Vavilova}(2013{\natexlab{a}})}]{Babyk:13a}
{Babyk}, I., \& {Vavilova}, I. 2013{\natexlab{a}}, ArXiv e-prints.
\newblock \doarXiv{1302.0873}

\bibitem[{{Babyk} \& {Vavilova}(2014)}]{Babyk:14a}
---. 2014, \apss, 349, 415, \dodoi{10.1007/s10509-013-1630-z}

\bibitem[{{Babyk}(2012{\natexlab{b}})}]{Babyk:12a}
{Babyk}, I.~V. 2012{\natexlab{b}}, Bulletin Crimean Astrophysical Observatory,
  108, 87, \dodoi{10.3103/S0190271712010044}

\bibitem[{{Babyk}(2018)}]{Babyk:18aasp}
---. 2018, Advances in Astronomy and Space Physics, 8, 28,
  \dodoi{10.17721/2227-1481.8.28-33}

\bibitem[{{Babyk}(2019)}]{Babyk:19}
---. 2019, Advances in Astronomy and Space Physics, 9, 8,
  \dodoi{10.17721/2227-1481.9.8-13}

\bibitem[{{Babyk} {et~al.}(2018{\natexlab{a}}){Babyk}, {McNamara}, {Nulsen},
  {Hogan}, {Vantyghem}, {Russell}, {Pulido}, \& {Edge}}]{Babyk:17scal}
{Babyk}, I.~V., {McNamara}, B.~R., {Nulsen}, P.~E.~J., {et~al.}
  2018{\natexlab{a}}, \apj, 857, 32, \dodoi{10.3847/1538-4357/aab3c9}

\bibitem[{{Babyk} {et~al.}(2018{\natexlab{b}}){Babyk}, {McNamara}, {Nulsen},
  {Russell}, {Vantyghem}, {Hogan}, \& {Pulido}}]{Babyk:17ent}
---. 2018{\natexlab{b}}, \apj, 862, 39, \dodoi{10.3847/1538-4357/aacce5}

\bibitem[{{Babyk} {et~al.}(2019){Babyk}, {McNamara}, {Tamhane}, {Nulsen},
  {Russell}, \& {Edge}}]{Babyk:17prof}
{Babyk}, I.~V., {McNamara}, B.~R., {Tamhane}, P.~D., {et~al.} 2019, \apj, 887,
  149, \dodoi{10.3847/1538-4357/ab54ce}

\bibitem[{{Babyk} \& {Vavilova}(2012)}]{Babyk:12oap}
{Babyk}, I.~V., \& {Vavilova}, I.~B. 2012, Odessa Astronomical Publications,
  25, 119

\bibitem[{{Babyk} \& {Vavilova}(2013{\natexlab{b}})}]{Babyk:13oap}
---. 2013{\natexlab{b}}, Odessa Astronomical Publications, 26, 175

\bibitem[{{Borgani} \& {Kravtsov}(2011)}]{Borgani:11}
{Borgani}, S., \& {Kravtsov}, A. 2011, Advanced Science Letters, 4, 204,
  \dodoi{10.1166/asl.2011.1209}

\bibitem[{{Borgani} {et~al.}(2004){Borgani}, {Murante}, {Springel}, {Diaferio},
  {Dolag}, {Moscardini}, {Tormen}, {Tornatore}, \& {Tozzi}}]{Borgani:04}
{Borgani}, S., {Murante}, G., {Springel}, V., {et~al.} 2004, \mnras, 348, 1078,
  \dodoi{10.1111/j.1365-2966.2004.07431.x}

\bibitem[{{Boroson} {et~al.}(2011){Boroson}, {Kim}, \& {Fabbiano}}]{Boroson:10}
{Boroson}, B., {Kim}, D.-W., \& {Fabbiano}, G. 2011, \apj, 729, 12,
  \dodoi{10.1088/0004-637X/729/1/12}

\bibitem[{{Burke} {et~al.}(2020){Burke}, {Laurino}, {Wmclaugh}, {Dtnguyen2},
  {Marie-Terrell}, {G{\"u}nther}, {Budynkiewicz}, {Siemiginowska}, {Aldcroft},
  {Deil}, {Sip{\H{o}}cz}, {Leinweber}, \& {Todd}}]{Burke:2020}
{Burke}, D., {Laurino}, O., {Wmclaugh}, {et~al.} 2020, {sherpa/sherpa: Sherpa
  4.12.1}, 4.12.1, Zenodo,  Zenodo, \dodoi{10.5281/zenodo.3944985}

\bibitem[{{Capelo} {et~al.}(2012){Capelo}, {Coppi}, \& {Natarajan}}]{Capelo:12}
{Capelo}, P.~R., {Coppi}, P.~S., \& {Natarajan}, P. 2012, \mnras, 422, 686,
  \dodoi{10.1111/j.1365-2966.2012.20648.x}

\bibitem[{{Chen} {et~al.}(2019){Chen}, {Avestruz}, {Kravtsov}, {Lau}, \&
  {Nagai}}]{Chen:2019}
{Chen}, H., {Avestruz}, C., {Kravtsov}, A.~V., {Lau}, E.~T., \& {Nagai}, D.
  2019, \mnras, 490, 2380, \dodoi{10.1093/mnras/stz2776}

\bibitem[{{Choi} {et~al.}(2015){Choi}, {Ostriker}, {Naab}, {Oser}, \&
  {Moster}}]{Choi:15}
{Choi}, E., {Ostriker}, J.~P., {Naab}, T., {Oser}, L., \& {Moster}, B.~P. 2015,
  \mnras, 449, 4105, \dodoi{10.1093/mnras/stv575}

\bibitem[{{Chon} {et~al.}(2016){Chon}, {Puchwein}, \&
  {B{\"o}hringer}}]{Chon:2016}
{Chon}, G., {Puchwein}, E., \& {B{\"o}hringer}, H. 2016, \aap, 592, A46,
  \dodoi{10.1051/0004-6361/201628532}

\bibitem[{{Dav{\'e}} {et~al.}(2008){Dav{\'e}}, {Oppenheimer}, \& {Sivanand
  am}}]{Dave:08}
{Dav{\'e}}, R., {Oppenheimer}, B.~D., \& {Sivanand am}, S. 2008, \mnras, 391,
  110, \dodoi{10.1111/j.1365-2966.2008.13906.x}

\bibitem[{{Del Popolo}(2002)}]{delpopolo:02}
{Del Popolo}, A. 2002, \mnras, 336, 81,
  \dodoi{10.1046/j.1365-8711.2002.05697.x}

\bibitem[{{Doe} {et~al.}(2007){Doe}, {Nguyen}, {Stawarz}, {Refsdal},
  {Siemiginowska}, {Burke}, {Evans}, {Evans}, {McDowell}, {Houck}, \&
  {Nowak}}]{Doe:2007}
{Doe}, S., {Nguyen}, D., {Stawarz}, C., {et~al.} 2007, in Astronomical Society
  of the Pacific Conference Series, Vol. 376, Astronomical Data Analysis
  Software and Systems XVI, ed. R.~A. {Shaw}, F.~{Hill}, \& D.~J. {Bell}, 543

\bibitem[{{Eckmiller} {et~al.}(2011){Eckmiller}, {Hudson}, \&
  {Reiprich}}]{Echmiller:11}
{Eckmiller}, H.~J., {Hudson}, D.~S., \& {Reiprich}, T.~H. 2011, \aap, 535,
  A105, \dodoi{10.1051/0004-6361/201116734}

\bibitem[{{Ettori}(2013)}]{Ettori:13}
{Ettori}, S. 2013, \mnras, 435, 1265, \dodoi{10.1093/mnras/stt1368}

\bibitem[{{Evrard} {et~al.}(2002){Evrard}, {MacFarland}, {Couchman}, {Colberg},
  {Yoshida}, {White}, {Jenkins}, {Frenk}, {Pearce}, {Peacock}, \&
  {Thomas}}]{Evrard:02}
{Evrard}, A.~E., {MacFarland}, T.~J., {Couchman}, H.~M.~P., {et~al.} 2002,
  \apj, 573, 7, \dodoi{10.1086/340551}

\bibitem[{{Farahi} {et~al.}(2018){Farahi}, {Guglielmo}, {Evrard}, {Poggianti},
  {Adami}, {Ettori}, {Gastaldello}, {Giles}, {Maughan}, {Rapetti}, {Sereno},
  {Altieri}, {Baldry}, {Birkinshaw}, {Bolzonella}, {Bongiorno}, {Brown},
  {Chiappetti}, {Driver}, {Elyiv}, {Garilli}, {Guennou}, {Hopkins}, {Iovino},
  {Koulouridis}, {Liske}, {Maurogordato}, {Owers}, {Pacaud}, {Pierre},
  {Plionis}, {Ponman}, {Robotham}, {Sadibekova}, {Scodeggio}, {Tuffs}, \&
  {Valtchanov}}]{Farahi:18}
{Farahi}, A., {Guglielmo}, V., {Evrard}, A.~E., {et~al.} 2018, \aap, 620, A8,
  \dodoi{10.1051/0004-6361/201731321}

\bibitem[{{Finoguenov} {et~al.}(2001){Finoguenov}, {Reiprich}, \&
  {B{\"o}hringer}}]{Finoguenov:01}
{Finoguenov}, A., {Reiprich}, T.~H., \& {B{\"o}hringer}, H. 2001, \aap, 368,
  749, \dodoi{10.1051/0004-6361:20010080}

\bibitem[{{Freeman} {et~al.}(2001){Freeman}, {Doe}, \&
  {Siemiginowska}}]{Freeman:2001}
{Freeman}, P., {Doe}, S., \& {Siemiginowska}, A. 2001, in Society of
  Photo-Optical Instrumentation Engineers (SPIE) Conference Series, Vol. 4477,
  Astronomical Data Analysis, ed. J.-L. {Starck} \& F.~D. {Murtagh}, 76--87,
  \dodoi{10.1117/12.447161}

\bibitem[{{Fruscione} {et~al.}(2006){Fruscione}, {McDowell}, {Allen},
  {Brickhouse}, {Burke}, {Davis}, {Durham}, {Elvis}, {Galle}, {Harris},
  {Huenemoerder}, {Houck}, {Ishibashi}, {Karovska}, {Nicastro}, {Noble},
  {Nowak}, {Primini}, {Siemiginowska}, {Smith}, \& {Wise}}]{Fruscione:06}
{Fruscione}, A., {McDowell}, J.~C., {Allen}, G.~E., {et~al.} 2006, in
  \procspie, Vol. 6270, Society of Photo-Optical Instrumentation Engineers
  (SPIE) Conference Series, 62701V, \dodoi{10.1117/12.671760}

\bibitem[{{Germain} {et~al.}(2006){Germain}, {Milaszewski}, {McLaughlin},
  {Miller}, {Evans}, {Evans}, \& {Burke}}]{Germain:2006}
{Germain}, G., {Milaszewski}, R., {McLaughlin}, W., {et~al.} 2006, in
  Astronomical Society of the Pacific Conference Series, Vol. 351, Astronomical
  Data Analysis Software and Systems XV, ed. C.~{Gabriel}, C.~{Arviset},
  D.~{Ponz}, \& S.~{Enrique}, 57

\bibitem[{{Giodini} {et~al.}(2013){Giodini}, {Lovisari}, {Pointecouteau},
  {Ettori}, {Reiprich}, \& {Hoekstra}}]{Giodini:13}
{Giodini}, S., {Lovisari}, L., {Pointecouteau}, E., {et~al.} 2013, \ssr, 177,
  247, \dodoi{10.1007/s11214-013-9994-5}

\bibitem[{{Hammami} \& {Mota}(2017)}]{Hammami:17}
{Hammami}, A., \& {Mota}, D.~F. 2017, \aap, 598, A132,
  \dodoi{10.1051/0004-6361/201629003}

\bibitem[{{Harris} {et~al.}(2017){Harris}, {Blakeslee}, \&
  {Harris}}]{Harris:17}
{Harris}, W.~E., {Blakeslee}, J.~P., \& {Harris}, G. L.~H. 2017, \apj, 836, 67,
  \dodoi{10.3847/1538-4357/836/1/67}

\bibitem[{{Hoekstra}(2007)}]{Hoekstra:07}
{Hoekstra}, H. 2007, \mnras, 379, 317, \dodoi{10.1111/j.1365-2966.2007.11951.x}

\bibitem[{{Hoekstra} {et~al.}(2013){Hoekstra}, {Bartelmann}, {Dahle}, {Israel},
  {Limousin}, \& {Meneghetti}}]{Hoekstra:13}
{Hoekstra}, H., {Bartelmann}, M., {Dahle}, H., {et~al.} 2013, \ssr, 177, 75,
  \dodoi{10.1007/s11214-013-9978-5}

\bibitem[{{Hogan} {et~al.}(2017){Hogan}, {McNamara}, {Pulido}, {Nulsen},
  {Vantyghem}, {Russell}, {Edge}, {Babyk}, {Main}, \& {McDonald}}]{Hogan:17a}
{Hogan}, M.~T., {McNamara}, B.~R., {Pulido}, F.~A., {et~al.} 2017, \apj, 851,
  66, \dodoi{10.3847/1538-4357/aa9af3}

\bibitem[{{Hu} \& {Kravtsov}(2003)}]{Hu:03}
{Hu}, W., \& {Kravtsov}, A.~V. 2003, \apj, 584, 702, \dodoi{10.1086/345846}

\bibitem[{{Kaiser}(1986)}]{Kaiser:86}
{Kaiser}, N. 1986, \mnras, 222, 323, \dodoi{10.1093/mnras/222.2.323}

\bibitem[{{Kaiser}(1991)}]{Kaiser:91}
---. 1991, \apj, 383, 104, \dodoi{10.1086/170768}

\bibitem[{{Kelly}(2007)}]{Kelly:07}
{Kelly}, B.~C. 2007, \apj, 665, 1489, \dodoi{10.1086/519947}

\bibitem[{{Kettula} {et~al.}(2013){Kettula}, {Finoguenov}, {Massey}, {Rhodes},
  {Hoekstra}, {Taylor}, {Spinelli}, {Tanaka}, {Ilbert}, {Capak}, {McCracken},
  \& {Koekemoer}}]{Kettula:13}
{Kettula}, K., {Finoguenov}, A., {Massey}, R., {et~al.} 2013, \apj, 778, 74,
  \dodoi{10.1088/0004-637X/778/1/74}

\bibitem[{{Kettula} {et~al.}(2015){Kettula}, {Giodini}, {van Uitert},
  {Hoekstra}, {Finoguenov}, {Lerchster}, {Erben}, {Heymans}, {Hildebrandt}, \&
  {Kitching}}]{Kettula:15}
{Kettula}, K., {Giodini}, S., {van Uitert}, E., {et~al.} 2015, \mnras, 451,
  1460, \dodoi{10.1093/mnras/stv923}

\bibitem[{{Kim} \& {Fabbiano}(2013)}]{Kim:13}
{Kim}, D.-W., \& {Fabbiano}, G. 2013, \apj, 776, 116,
  \dodoi{10.1088/0004-637X/776/2/116}

\bibitem[{{Kim} \& {Fabbiano}(2015)}]{Kim:15}
---. 2015, \apj, 812, 127, \dodoi{10.1088/0004-637X/812/2/127}

\bibitem[{{Kravtsov} {et~al.}(2006){Kravtsov}, {Vikhlinin}, \&
  {Nagai}}]{Kravtsov:06}
{Kravtsov}, A.~V., {Vikhlinin}, A., \& {Nagai}, D. 2006, \apj, 650, 128,
  \dodoi{10.1086/506319}

\bibitem[{{Kubo} {et~al.}(2009){Kubo}, {Annis}, {Hardin}, {Kubik}, {Lawhorn},
  {Lin}, {Nicklaus}, {Nelson}, {Reis}, \& {Seo}}]{Kubo:09}
{Kubo}, J.~M., {Annis}, J., {Hardin}, F.~M., {et~al.} 2009, \apj, 702, L110,
  \dodoi{10.1088/0004-637X/702/2/L110}

\bibitem[{{Le Brun} {et~al.}(2014){Le Brun}, {McCarthy}, {Schaye}, \&
  {Ponman}}]{LeBrun2014}
{Le Brun}, A. M.~C., {McCarthy}, I.~G., {Schaye}, J., \& {Ponman}, T.~J. 2014,
  \mnras, 441, 1270, \dodoi{10.1093/mnras/stu608}

\bibitem[{{Li} {et~al.}(2015){Li}, {Bryan}, {Ruszkowski}, {Voit}, {O'Shea}, \&
  {Donahue}}]{Li:15}
{Li}, Y., {Bryan}, G.~L., {Ruszkowski}, M., {et~al.} 2015, \apj, 811, 73,
  \dodoi{10.1088/0004-637X/811/2/73}

\bibitem[{{Lieu} {et~al.}(2016){Lieu}, {Smith}, {Giles}, {Ziparo}, {Maughan},
  {D{\'e}mocl{\`e}s}, {Pacaud}, {Pierre}, {Adami}, {Bah{\'e}}, {Clerc},
  {Chiappetti}, {Eckert}, {Ettori}, {Lavoie}, {Le Fevre}, {McCarthy},
  {Kilbinger}, {Ponman}, {Sadibekova}, \& {Willis}}]{Lieu:16}
{Lieu}, M., {Smith}, G.~P., {Giles}, P.~A., {et~al.} 2016, \aap, 592, A4,
  \dodoi{10.1051/0004-6361/201526883}

\bibitem[{{Lin} \& {Mohr}(2004)}]{Lin:04}
{Lin}, Y.-T., \& {Mohr}, J.~J. 2004, \apj, 617, 879, \dodoi{10.1086/425412}

\bibitem[{{Lovisari} {et~al.}(2015){Lovisari}, {Reiprich}, \&
  {Schellenberger}}]{Lovisari:15}
{Lovisari}, L., {Reiprich}, T.~H., \& {Schellenberger}, G. 2015, \aap, 573,
  A118, \dodoi{10.1051/0004-6361/201423954}

\bibitem[{{Ma} {et~al.}(2013){Ma}, {McNamara}, \& {Nulsen}}]{Ma:13}
{Ma}, C.-J., {McNamara}, B.~R., \& {Nulsen}, P.~E.~J. 2013, \apj, 763, 63,
  \dodoi{10.1088/0004-637X/763/1/63}

\bibitem[{{Mahdavi} {et~al.}(2000){Mahdavi}, {B{\"o}hringer}, {Geller}, \&
  {Ramella}}]{Mahdavi:00}
{Mahdavi}, A., {B{\"o}hringer}, H., {Geller}, M.~J., \& {Ramella}, M. 2000,
  \apj, 534, 114, \dodoi{10.1086/308740}

\bibitem[{{Mahdavi} {et~al.}(2013){Mahdavi}, {Hoekstra}, {Babul}, {Bildfell},
  {Jeltema}, \& {Henry}}]{Mahdavi:13}
{Mahdavi}, A., {Hoekstra}, H., {Babul}, A., {et~al.} 2013, \apj, 767, 116,
  \dodoi{10.1088/0004-637X/767/2/116}

\bibitem[{{Main} {et~al.}(2017){Main}, {McNamara}, {Nulsen}, {Russell}, \&
  {Vantyghem}}]{Main:17}
{Main}, R.~A., {McNamara}, B.~R., {Nulsen}, P.~E.~J., {Russell}, H.~R., \&
  {Vantyghem}, A.~N. 2017, \mnras, 464, 4360, \dodoi{10.1093/mnras/stw2644}

\bibitem[{{Mantz} {et~al.}(2016){Mantz}, {Allen}, {Morris}, {von der Linden},
  {Applegate}, {Kelly}, {Burke}, {Donovan}, \& {Ebeling}}]{Mantz:16}
{Mantz}, A.~B., {Allen}, S.~W., {Morris}, R.~G., {et~al.} 2016, \mnras, 463,
  3582, \dodoi{10.1093/mnras/stw2250}

\bibitem[{{Markevitch}(1998)}]{Markevitch:98}
{Markevitch}, M. 1998, \apj, 504, 27, \dodoi{10.1086/306080}

\bibitem[{{Martino} {et~al.}(2014){Martino}, {Mazzotta}, {Bourdin}, {Smith},
  {Bartalucci}, {Marrone}, {Finoguenov}, \& {Okabe}}]{Martino:14}
{Martino}, R., {Mazzotta}, P., {Bourdin}, H., {et~al.} 2014, \mnras, 443, 2342,
  \dodoi{10.1093/mnras/stu1267}

\bibitem[{{McNamara} \& {Nulsen}(2007)}]{McNamara:07}
{McNamara}, B.~R., \& {Nulsen}, P.~E.~J. 2007, \araa, 45, 117,
  \dodoi{10.1146/annurev.astro.45.051806.110625}

\bibitem[{{McNamara} \& {Nulsen}(2012)}]{McNamara:12}
---. 2012, New Journal of Physics, 14, 055023,
  \dodoi{10.1088/1367-2630/14/5/055023}

\bibitem[{{Meneghetti} {et~al.}(2010){Meneghetti}, {Rasia}, {Merten},
  {Bellagamba}, {Ettori}, {Mazzotta}, {Dolag}, \& {Marri}}]{Meneghetti:10}
{Meneghetti}, M., {Rasia}, E., {Merten}, J., {et~al.} 2010, \aap, 514, A93,
  \dodoi{10.1051/0004-6361/200913222}

\bibitem[{{Morandi} \& {Ettori}(2007)}]{Morandi:07}
{Morandi}, A., \& {Ettori}, S. 2007, \mnras, 380, 1521,
  \dodoi{10.1111/j.1365-2966.2007.12158.x}

\bibitem[{{Mushotzky}(1984)}]{Mushotzky:84}
{Mushotzky}, R.~F. 1984, Physica Scripta Volume T, 7, 157,
  \dodoi{10.1088/0031-8949/1984/T7/036}

\bibitem[{{Mushotzky} \& {Scharf}(1997)}]{Mushotzky:97}
{Mushotzky}, R.~F., \& {Scharf}, C.~A. 1997, \apjl, 482, L13,
  \dodoi{10.1086/310676}

\bibitem[{{Nagai} {et~al.}(2007){Nagai}, {Vikhlinin}, \& {Kravtsov}}]{Nagai:07}
{Nagai}, D., {Vikhlinin}, A., \& {Kravtsov}, A.~V. 2007, \apj, 655, 98,
  \dodoi{10.1086/509868}

\bibitem[{{Nevalainen} {et~al.}(2000){Nevalainen}, {Markevitch}, \&
  {Forman}}]{Nevalainen:00}
{Nevalainen}, J., {Markevitch}, M., \& {Forman}, W. 2000, \apj, 532, 694,
  \dodoi{10.1086/308608}

\bibitem[{{O'Hara} {et~al.}(2006){O'Hara}, {Mohr}, {Bialek}, \&
  {Evrard}}]{Ohara:06}
{O'Hara}, T.~B., {Mohr}, J.~J., {Bialek}, J.~J., \& {Evrard}, A.~E. 2006, \apj,
  639, 64, \dodoi{10.1086/499327}

\bibitem[{{Okabe} \& {Smith}(2016)}]{Okabe:16}
{Okabe}, N., \& {Smith}, G.~P. 2016, \mnras, 461, 3794,
  \dodoi{10.1093/mnras/stw1539}

\bibitem[{{O'Sullivan} {et~al.}(2003){O'Sullivan}, {Ponman}, \&
  {Collins}}]{OSullivan_sample:03}
{O'Sullivan}, E., {Ponman}, T.~J., \& {Collins}, R.~S. 2003, \mnras, 340, 1375,
  \dodoi{10.1046/j.1365-8711.2003.06396.x}

\bibitem[{{Pellegrini} {et~al.}(2018){Pellegrini}, {Ciotti}, {Negri}, \&
  {Ostriker}}]{Pellegrini2018}
{Pellegrini}, S., {Ciotti}, L., {Negri}, A., \& {Ostriker}, J.~P. 2018, \apj,
  856, 115, \dodoi{10.3847/1538-4357/aaae07}

\bibitem[{{Pike} {et~al.}(2014){Pike}, {Kay}, {Newton}, {Thomas}, \&
  {Jenkins}}]{Pike2014}
{Pike}, S.~R., {Kay}, S.~T., {Newton}, R. D.~A., {Thomas}, P.~A., \& {Jenkins},
  A. 2014, \mnras, 445, 1774, \dodoi{10.1093/mnras/stu1788}

\bibitem[{{Planelles} {et~al.}(2015){Planelles}, {Schleicher}, \&
  {Bykov}}]{Planelles:15}
{Planelles}, S., {Schleicher}, D.~R.~G., \& {Bykov}, A.~M. 2015, \ssr, 188, 93,
  \dodoi{10.1007/s11214-014-0045-7}

\bibitem[{{Ponman} {et~al.}(1996){Ponman}, {Bourner}, {Ebeling}, \&
  {B{\"o}hringer}}]{Ponman:96}
{Ponman}, T.~J., {Bourner}, P.~D.~J., {Ebeling}, H., \& {B{\"o}hringer}, H.
  1996, \mnras, 283, 690, \dodoi{10.1093/mnras/283.2.690}

\bibitem[{{Pulatova} {et~al.}(2015){Pulatova}, {Vavilova}, {Sawangwit},
  {Babyk}, \& {Klimanov}}]{Pulatova:15}
{Pulatova}, N.~G., {Vavilova}, I.~B., {Sawangwit}, U., {Babyk}, I., \&
  {Klimanov}, S. 2015, \mnras, 447, 2209, \dodoi{10.1093/mnras/stu2556}

\bibitem[{{Pulido} {et~al.}(2018){Pulido}, {McNamara}, {Edge}, {Hogan},
  {Vantyghem}, {Russell}, {Nulsen}, {Babyk}, \& {Salom{\'e}}}]{Pulido:17}
{Pulido}, F.~A., {McNamara}, B.~R., {Edge}, A.~C., {et~al.} 2018, \apj, 853,
  177, \dodoi{10.3847/1538-4357/aaa54b}

\bibitem[{{Rasia} {et~al.}(2005){Rasia}, {Mazzotta}, {Borgani}, {Moscardini},
  {Dolag}, {Tormen}, {Diaferio}, \& {Murante}}]{Rasia:05}
{Rasia}, E., {Mazzotta}, P., {Borgani}, S., {et~al.} 2005, \apjl, 618, L1,
  \dodoi{10.1086/427554}

\bibitem[{{Rasia} {et~al.}(2006){Rasia}, {Ettori}, {Moscardini}, {Mazzotta},
  {Borgani}, {Dolag}, {Tormen}, {Cheng}, \& {Diaferio}}]{Rasia:06}
{Rasia}, E., {Ettori}, S., {Moscardini}, L., {et~al.} 2006, \mnras, 369, 2013,
  \dodoi{10.1111/j.1365-2966.2006.10466.x}

\bibitem[{{Rosito} {et~al.}(2021){Rosito}, {Pedrosa}, {Tissera}, {Chisari},
  {Dom{\'\i}nguez-Tenreiro}, {Dubois}, {Peirani}, {Devriendt}, {Pichon}, \&
  {Slyz}}]{Rosito:2021}
{Rosito}, M.~S., {Pedrosa}, S.~E., {Tissera}, P.~B., {et~al.} 2021, \aap, 652,
  A44, \dodoi{10.1051/0004-6361/202039976}

\bibitem[{{Shimizu} {et~al.}(2003){Shimizu}, {Kitayama}, {Sasaki}, \&
  {Suto}}]{Shimizu:03}
{Shimizu}, M., {Kitayama}, T., {Sasaki}, S., \& {Suto}, Y. 2003, \apj, 590,
  197, \dodoi{10.1086/367955}

\bibitem[{{Sijacki} \& {Springel}(2006)}]{Sijacki:06}
{Sijacki}, D., \& {Springel}, V. 2006, \mnras, 371, 1025,
  \dodoi{10.1111/j.1365-2966.2006.10752.x}

\bibitem[{{Smith} {et~al.}(2016){Smith}, {Mazzotta}, {Okabe}, {Ziparo},
  {Mulroy}, {Babul}, {Finoguenov}, {McCarthy}, {Lieu}, {Bah{\'e}}, {Bourdin},
  {Evrard}, {Futamase}, {Haines}, {Jauzac}, {Marrone}, {Martino}, {May},
  {Taylor}, \& {Umetsu}}]{Smith:16}
{Smith}, G.~P., {Mazzotta}, P., {Okabe}, N., {et~al.} 2016, \mnras, 456, L74,
  \dodoi{10.1093/mnrasl/slv175}

\bibitem[{{Sun} {et~al.}(2009){Sun}, {Voit}, {Donahue}, {Jones}, {Forman}, \&
  {Vikhlinin}}]{Sun:09}
{Sun}, M., {Voit}, G.~M., {Donahue}, M., {et~al.} 2009, \apj, 693, 1142,
  \dodoi{10.1088/0004-637X/693/2/1142}

\bibitem[{{Tozzi} {et~al.}(2000){Tozzi}, {Scharf}, \& {Norman}}]{Tozzi:00}
{Tozzi}, P., {Scharf}, C., \& {Norman}, C. 2000, \apj, 542, 106,
  \dodoi{10.1086/309500}

\bibitem[{{Vavilova} \& {Babyk}(2018)}]{Babyk:18oap}
{Vavilova}, I.~B., \& {Babyk}, I.~V. 2018, Odessa Astronomical Publications,
  31, 239, \dodoi{10.18524/1810-4215.2018.31.146678}

\bibitem[{{Vavilova} {et~al.}(2015){Vavilova}, {Bolotin}, {Boyarsky},
  {Danevich}, {Kobychev}, {Tretyak}, {Babyk}, {Iakubovskyi}, {Hnatyk}, \&
  {Sergeev}}]{Babyk_book:15}
{Vavilova}, I.~B., {Bolotin}, Y.~L., {Boyarsky}, A.~M., {et~al.} 2015, {Dark
  matter: Observational manifestation and experimental searches}

\bibitem[{{Vikhlinin} {et~al.}(2006){Vikhlinin}, {Kravtsov}, {Forman}, {Jones},
  {Markevitch}, {Murray}, \& {Van Speybroeck}}]{Vikhlinin:06}
{Vikhlinin}, A., {Kravtsov}, A., {Forman}, W., {et~al.} 2006, \apj, 640, 691,
  \dodoi{10.1086/500288}

\bibitem[{{Vikhlinin} {et~al.}(2009){Vikhlinin}, {Burenin}, {Ebeling},
  {Forman}, {Hornstrup}, {Jones}, {Kravtsov}, {Murray}, {Nagai}, {Quintana}, \&
  {Voevodkin}}]{Vikhlinin:09}
{Vikhlinin}, A., {Burenin}, R.~A., {Ebeling}, H., {et~al.} 2009, \apj, 692,
  1033, \dodoi{10.1088/0004-637X/692/2/1033}

\bibitem[{{Voit}(2005)}]{Voit05}
{Voit}, G.~M. 2005, Reviews of Modern Physics, 77, 207,
  \dodoi{10.1103/RevModPhys.77.207}

\bibitem[{{Voit} {et~al.}(2005){Voit}, {Kay}, \& {Bryan}}]{Voit:05}
{Voit}, G.~M., {Kay}, S.~T., \& {Bryan}, G.~L. 2005, \mnras, 364, 909,
  \dodoi{10.1111/j.1365-2966.2005.09621.x}

\bibitem[{{Xu} {et~al.}(2001){Xu}, {Jin}, \& {Wu}}]{Xu:01}
{Xu}, H., {Jin}, G., \& {Wu}, X.-P. 2001, \apj, 553, 78, \dodoi{10.1086/320662}

\bibitem[{{Xue} \& {Wu}(2000)}]{Xue:00}
{Xue}, Y.-J., \& {Wu}, X.-P. 2000, \mnras, 318, 715,
  \dodoi{10.1046/j.1365-8711.2000.03753.x}

\bibitem[{{Zhang} {et~al.}(2008){Zhang}, {Finoguenov}, {B{\"o}hringer},
  {Kneib}, {Smith}, {Kneissl}, {Okabe}, \& {Dahle}}]{Zhang:08}
{Zhang}, Y.~Y., {Finoguenov}, A., {B{\"o}hringer}, H., {et~al.} 2008, \aap,
  482, 451, \dodoi{10.1051/0004-6361:20079103}

\end{thebibliography}
\bibliographystyle{aasjournal}



\end{document}